\documentclass[a4paper,11pt]{article}
\usepackage{graphicx}
\usepackage{epsf,amsmath,bbold,amsfonts,stmaryrd}

\usepackage[utf8]{inputenc}
\usepackage{mathrsfs}
\usepackage{appendix}
\usepackage{amssymb}
\usepackage{float}
\usepackage{color}
\usepackage{cite}
\usepackage{hyperref}
\hypersetup{pageanchor=false}
\usepackage{indentfirst}
\usepackage{url}
\usepackage{float}
\usepackage{caption}
\usepackage[numbers,square,comma,sort&compress,merge]{natbib}
\usepackage{graphicx}
\usepackage{animate}
\usepackage{subfigure}

\def\b{\beta}

\def\D{\Delta}

\def\hs{\hspace}

\newcommand{\bra}{\left\langle}
\newcommand{\ket}{\right\rangle}

\def\nn{\nonumber}

\def\tmu{\tilde \mu}

\def\be{\begin{equation}}
\def\ee{\end{equation}}

\def\bea{\begin{eqnarray}}
\def\eea{\end{eqnarray}}

\def\ba{\begin{array}}
\def\ea{\end{array}}

\def\bc{\begin{center}}
\def\ec{\end{center}}

\def\bl{\begin{flushleft}}
\def\el{\end{flushleft}}

\def\br{\begin{flushright}}
\def\er{\end{flushright}}

\def\bi{\begin{itemize}}
\def\ei{\end{itemize}}

\def\bt{\begin{tabular}}
\def\et{\end{tabular}}

\numberwithin{equation}{section}

\begin{document}

\title{\textbf{Entanglement Entropy in $T\overline{T}$-Deformed CFT}}
\author{Bin Chen$^{1,2,3}$, Lin Chen$^{1}$ and Peng-xiang Hao$^{1}$\thanks{bchen01@pku.edu.cn, linchen91@pku.edu.cn, pxhao@pku.edu.cn}}
\maketitle
\begin{center}
\textsl{$^{1}$Department of Physics and State Key Laboratory of Nuclear
Physics and Technology, Peking University, No.5 Yiheyuan Rd, Beijing
100871, P.R. China}\\
\textsl{$^{1}$Center for High Energy Physics, Peking University,
No.5 Yiheyuan Rd, Beijing 100871, P. R. China}\\
\textsl{$^{3}$Collaborative Innovation Center of Quantum Matter,
No.5 Yiheyuan Rd, Beijing 100871, P. R. China}\\
\par\end{center}

\date{}

\maketitle

\vspace{-10mm}

\vspace{8mm}

\begin{abstract}
In this paper, we study the entanglement entropy of a single interval on a cylinder in  two-dimensional $T\overline{T}$-deformed conformal field theory. For such case, the (R\'enyi) entanglement entropy takes a universal form in a CFT. We compute the correction due to the deformation up to the leading order of the deformation parameter in the framework of the conformal perturbation theory. We find that the correction to the entanglement entropy is nonvanishing in the finite temperature case, while it is vanishing in the finite size case. For the deformed holographic large $c$ CFT, which is proposed to be dual to a AdS$_3$ gravity in a finite region, we find the agreement with the holographic entanglement entropy via the Ryu-Takayanagi formula. Moreover, we compute the leading order correction to the R\'enyi entropy, and discuss its holographic picture as well.

\end{abstract}
\baselineskip 18pt

\thispagestyle{empty}
\newpage

\tableofcontents

\section{Introduction}

The integrable quantum field theory allows us to understand the non-perturbative aspects of the quantum field theory. In a remarkable paper by Zamolodchikov\cite{ttcft}, the operator $T_{zz}T_{\bar z \bar z}-T^2_{z\bar z}$ of a two-dimensional(2D) quantum field theory(QFT) was studied and its expectation value  
had been shown to have an analytic form. 
 Such deformation is now called $T\overline{T}$-deformation. The $T\overline{T}$ deformation has some interesting properties, as shown in various studies including the spectrum and the S-matrix\cite{ttcft2}. In particular, an integrable QFT deformed by such operator was found to be still integrable \cite{ttcft2,ttcft3}. In \cite{ttcft3}, it was shown that the deformation of the theory of 24 free scalars leads to the Nambu-Goto action. In \cite{cardytt}, Cardy explained the solvability of the deformation by considering it as a stochastic process.
For other studies on the $T\overline{T}$ deformation of a field theory, see \cite{ttatlargec,ttinclosedform,ttpartition}.

The $T\overline{T}$ deformation of a 2D conformal field theory(CFT) is of particular interest. To be more precise
the $T\overline{T}$ deformed CFT form a one-parameter family of the theories $\mathcal{T}^{(\mu)}$ parametrized by $\mu\geq 0$. The original CFT sits on $\mu=0$. Moving infinitesimally from $\mathcal{T}^{(\mu)}$ to $\mathcal{T}^{(\mu+\delta \mu)}$ is achieved by adding a term
\begin{eqnarray}\label{ttdeform}
\delta \mu \int d^2x \left(T^{(\mu)}\overline{T}^{(\mu)}-\Theta^{(\mu)2}\right)
\end{eqnarray}
to the action of $\mathcal{T}^{(\mu)}$, where
\begin{eqnarray}
T^{(\mu)}=-2\pi T_{zz}^{(\mu)},\;\;\overline{T}^{(\mu)}=-2\pi T_{\bar{z}\bar{z}}^{(\mu)},\;\;\Theta^{(\mu)}=2\pi T_{z\bar{z}}^{(\mu)}
\end{eqnarray}
are the stress tensor of $\mathcal{T}^{(\mu)}$. 
In this case, the spectrum could be determined explicitly. Considering the deformed CFT  on a cylinder of circumference $L$, the spectrum is
 \be
 E_n(\mu,L)L=\frac{2\pi}{\tmu}\left(1-\sqrt{1-2\tmu M_n+\tmu^2J^2_n} \right), \label{spectrum}
 \ee
 where $ \tmu=\frac{\mu}{4\pi L^2}$ is a dimensionless quantity and
 \be
M_n=\D_n+\bar\D_n-\frac{c}{12}, \hs{3ex}J_n=\D_n-\bar\D_n
 \ee
 are the conformal dimensions and the spins of the primary operators in the undeformed CFT.
 As the spectrum could be imaginary for a fixed $\tmu$, the theory should  have a UV cutoff.  It is certainly an interesting problem to find a UV completion of such deformation.

On the other hand, the $T\overline{T}$-deformation opens a new window to study the AdS/CFT correspondence. It is a double-trace deformation, and could change the boundary condition of the AdS gravity. For a $T\overline{T}$-deformed holographic CFT, McGough, Mezei, and Verlinde \cite{verlinde} proposed that the dual AdS$_3$ gravity should be defined in a finite region, with the asymptotic boundary being at a finite radial position.
More precisely if a CFT i.e. $\mathcal{T}^{(0)}$ has a gravity dual, then the theory $\mathcal{T}^{(\mu)}$ is dual to the original gravitational theory with the new boundary at $r=r_c$. With our convention, the relation between $\mu$ and $r_c$ is
\begin{eqnarray}\label{muandrc}
\mu=\frac{6R^4}{\pi c r_c^2},
\end{eqnarray}
where $R$ is the AdS radius, and $c$ is the central charge of the original CFT. This new correspondence has been checked from various points of view. First of all, the spectrum (\ref{spectrum}) is reproduced by considering the quasi-local energy of a BTZ black hole of mass $M_n$ and angular momentum $J_n$ in a spatial region $r<r_c$. Secondly, the superluminal propagation of the perturbation of the stress tensor\cite{Cardy:2015xaa} can be understood holographically by the metric perturbations preserving Dirichlet boundary condition on the surface $r=r_c$\cite{Marolf:2012dr}. Moreover, the exact RG equation could be understood holographically as well\cite{verlinde}. More on the holographic interpretation of the $T\overline{T}$ deformation can be found in \cite{kutasov1,kutasov2,dubovsky1,dubovsky2,bihdtott,ttandcf,cutoffads,commentsontt,ttingenerald}\footnote{There is another interesting generalization of the $T\overline{T}$ deformation, the so-called $T\overline{J}$-deformation\cite{Guica:2017lia}, which breaks the Lorentz symmetry but is still solvable. For other relevant studies on this kind of deformation, see \cite{Bzowski:2018pcy,Chakraborty:2018vja,Apolo:2018qpq}.}.

In this paper, we would like to study the entanglement entropy in the $T\overline{T}$-deformed conformal field theory. In particular we pay special attention to the entanglement entropy in the deformed holographic CFT and investigate its implication in the AdS/CFT correspondence. In a holographic CFT, the entanglement entropy could be captured by the area of the minimal surface via the Ryu-Takayanagi(RT) formula \cite{rtformula,Ryu:2006ef}. When considering the new duality proposed in \cite{verlinde}, it seems that the RT-formula still holds. We would like to use the entanglement entropy to test their proposal. More concretely we are going to compute  the entanglement entropy of a single interval on  a cylinder in the $T\overline{T}$-deformed CFT by using  the conformal perturbation method. We will investigate two cases: the one at a finite temperature and the other one with a finite size. We find that in the finite temperature case, there is indeed nonvanishing correction from the deformation, while in the finite size case, the correction is vanishing. We discuss the holographic entanglement entropy via the RT formula and find the consistent picture. Moreover we compute analytically the leading order correction to the R\'enyi entropy, and discuss its holographic picture. We show that for the AdS$_3$ gravity with a cutoff surface, the on-shell action includes a cutoff-dependent term, which corresponds to the leading order correction due to the $T\overline{T}$-deformation in the partition function in the CFT. 

The remaining parts of the paper are organized as follows. In section 2, we compute perturbatively the single-interval (Renyi) entanglement entropy on a cylinder in the deformed CFT. In section 3, we compute the holographic entanglement entropy of a single interval in the BTZ black hole and global AdS$_3$ with a finite radius cutoff, and compare with the field theory results. In section 4, we show that the on-shell action of the gravitational configuration in a cut-off restrained region could be dual to the CFT partition function with the leading order correction under the $T\overline{T}$-deformation. We end with discussions in section 5. In the appendix, we collect some technical details.

While this papar was in preparation, closely related studies were presented in \cite{eeandtt}. The authors in \cite{eeandtt}  considered the entanglement entropy for an entangling surface consisting of two antipodal points on a sphere. 

\section{Entanglement entropy in $T\overline{T}$-deformed CFT}
Let us consider a $T\overline{T}$-deformed CFT living on some manifold $\mathcal{M}$. And we are interested in the entanglement entropy of some subsystem $A \in \mathcal{M}$. The entanglement entropy is given by
\begin{eqnarray}
S(A)=\lim_{n\to1}S_n(A),\;\;\;\;S_n(A)=\frac{1}{1-n}\log\frac{Z_n(A)}{Z^n},
\end{eqnarray}
where $Z$ is the partition function on $\mathcal{M}$, $Z_n(A)$ is the partition function on the manifold $\mathcal{M}^n(A)$ which is obtained by gluing $n$ copies of $\mathcal{M}$ together along $A$. The precise definition of $\mathcal{M}^n$ and more details about the above formulas can be found in \cite{cardy}. In this work, we only consider the small $\mu$ case, i.e. $\mu\to0$. According to (\ref{ttdeform}) the action of the deformed CFT can be written as
\begin{eqnarray}
S=S_{CFT}+\mu \int_{\mathcal{M}}(T\overline{T}-\Theta^2),
\end{eqnarray}
where $T,\overline{T}$ and $\Theta$ are the quantities of the original CFT. Now we have
\begin{eqnarray}
\frac{Z_n(A)}{Z^n}=\frac{\int_{\mathcal{M}^n} e^{-S_{CFT}-\mu \int_{\mathcal{M}^n}(T\overline{T}-\Theta^2)}}{\left[\int_{\mathcal{M}} e^{-S_{CFT}-\mu \int_{\mathcal{M}}(T\overline{T}-\Theta^2)}\right]^n}.
\end{eqnarray}
Since $\mu$ is small, we further expand in terms of $\mu$ and get
\begin{eqnarray}
\frac{Z_n(A)}{Z^n}=\frac{\int_{\mathcal{M}^n} e^{-S_{CFT}}\left(1-\mu\int_{\mathcal{M}^n}(T\overline{T}-\Theta^2)+O(\mu^2)\right)}{\left[\int_{\mathcal{M}} e^{-S_{CFT}}\left(1-\mu\int_{\mathcal{M}}(T\overline{T}-\Theta^2)+O(\mu^2)\right)\right]^n}.
\end{eqnarray}

We know that in a CFT which is defined on a flat manifold, any correlation function with $T^\mu_\mu$ insertion is zero, i.e. $\left<T^\mu_\mu\dots\right>=0$. Later we will always consider the case $\mathcal{M}$ is a cylinder. Thus
\begin{eqnarray}
\int_{\mathcal{M}} e^{-S_{CFT}}\Theta^2&\sim&\left<\Theta^2\right>_{\mathcal{M}}=0,\\
\int_{\mathcal{M}^n} e^{-S_{CFT}}\Theta^2&\sim&\left<\Theta^2\sigma\right>_{\mathcal{M}}=0,
\end{eqnarray}
with $\sigma$  being the operator inducing the field identification such that the adjacent replicas are pasted along $A$.   After some simple algebra, we get
\begin{eqnarray}
\frac{Z_n(A)}{Z^n}=\left(\frac{\int_{\mathcal{M}^n} e^{-S_{CFT}}}{\left[\int_{\mathcal{M}} e^{-S_{CFT}}\right]^n}\right)\left(1-\mu\int_{\mathcal{M}^n}\left<T\overline{T}\right>_{\mathcal{M}^n}+n\mu\int_{\mathcal{M}}\left<T\overline{T}\right>_{\mathcal{M}}+O(\mu^2)\right).
\end{eqnarray}

Notice that $\left<T\overline{T}\right>_{\mathcal{M}^n}$ is only a function defined on $\mathcal{M}^n$. Actually we have
\begin{eqnarray}
\int_{\mathcal{M}^n}\left<T\overline{T}\right>_{\mathcal{M}^n}=n\int_{\mathcal{M}}\left<T\overline{T}\right>_{\mathcal{M}^n},
\end{eqnarray}
from which we get
\begin{eqnarray}
\frac{Z_n(A)}{Z^n}=\left(\frac{\int_{\mathcal{M}^n} e^{-S_{CFT}}}{\left[\int_{\mathcal{M}} e^{-S_{CFT}}\right]^n}\right)\left(1-n\mu\int_{\mathcal{M}}\left[\left<T\overline{T}\right>_{\mathcal{M}^n}-\left<T\overline{T}\right>_{\mathcal{M}}\right]+O(\mu^2)\right).
\end{eqnarray}
Then we can read the leading order correction to $S_n(A)$
\begin{eqnarray}\label{deltasn}
\delta S_n(A)=\frac{-n\mu}{1-n}\int_{\mathcal{M}}\left[\left<T\overline{T}\right>_{\mathcal{M}^n}-\left<T\overline{T}\right>_{\mathcal{M}}\right].
\end{eqnarray}
Taking the $n\to1$ limit, we have  the leading order correction to $S(A)$. In the following, let us consider two concrete cases where $\delta S(A)$ can be calculated.

\subsection{Finite temperature}\label{finitet}

The first case is  a $2D$ deformed CFT at a finite temperature $1/\beta$. The spatial direction is not compactified and the manifold $\mathcal{M}$ on which the theory is defined is an infinitely long cylinder with circumference $\beta$. We introduce complex coordinate $w=x+i\tau$ and $\bar{w}=x-i\tau$ on the cylinder $\mathcal{M}$, where $x\in(-\infty,\infty)$ and $\tau\in(0,\beta)$ with the identification $\tau\sim\tau+\beta$.  The subsystem $A$ is chosen to be a single interval of length $l$ which will be parallel to the axis of the cylinder.
The endpoints of $A$ are put at $(w,\bar{w})=(0,0)$ and $(w,\bar{w})=(l,l)$.

Consider the transformation
\begin{eqnarray}
w\to z=e^{\frac{2\pi w}{\beta}},
\end{eqnarray}
which maps the cylinder to a plane $\mathcal{C}$. The  stress tensor obeys the well-known transformation law
\begin{eqnarray}\label{sttransf}
T(w)=\left(\frac{dz}{dw}\right)^2T(z)+\frac{c}{12}\{z,w\},
\end{eqnarray}
where
\be
\{z,w\}=(z^{\prime\prime\prime}z^{\prime}-\frac{3}{2}z^{\prime\prime2})/z^{\prime2}
\ee
 is the Schwarzian derivative. There is a similar relation for $\overline{T}$. Using (\ref{sttransf})  and $\left<T(z)\right>_{\mathcal{C}}=0$, we find
\begin{eqnarray}\label{tt1}
\nonumber
\left<T\overline{T}(w,\bar{w})\right>_{\mathcal{M}}&=&\left(\frac{c}{12}\right)^2\{z,w\}\{\bar{z},\bar{w}\}\\
&=&\left(\frac{c}{12}\right)^2\left(\frac{2\pi^2}{\beta^2}\right)^2.
\end{eqnarray}

To obtain $\left<T\overline{T}(w,\bar{w})\right>_{\mathcal{M}^n}$, one should consider the following maps. The first map is
\begin{eqnarray}
w\to w^\prime=e^{\frac{2\pi w}{\beta}},
\end{eqnarray}
which maps each sheet of $\mathcal{M}^n$ to a plane $\mathcal{C}$. The interval $A$ on the cylinder $\mathcal{M}$ is mapped to an interval $A^\prime$ on the plane $\mathcal{C}$ whose endpoints become $(w^\prime,\bar{w}^\prime)=(1,1)$ and $(w^\prime,\bar{w}^\prime)=(e^{\frac{2\pi l}{\beta}},e^{\frac{2\pi l}{\beta}})$. After this map $\mathcal{M}^n$ becomes a manifold $\mathcal{C}^n$ which is obtained by gluing $n$ copies of the plane $\mathcal{C}$ together along $A^\prime$. The next map is
\begin{eqnarray}
w^\prime\to z=\left(\frac{w^\prime-1}{w^\prime-e^{\frac{2\pi l}{\beta}}}\right)^{\frac{1}{n}},
\end{eqnarray}
which maps $\mathcal{C}^n$ to a plane $\mathcal{C}$. More about this map can be found in section 3 of \cite{cardy}. Combining these two maps, we find a map $w\to z$ relating $\mathcal{M}^n$ to the plane $\mathcal{C}$.
Once again using (\ref{sttransf}) and $\left<T(z)\right>_{\mathcal{C}}=0$, we find
\begin{eqnarray}
\left<T\overline{T}(w,\bar{w})\right>_{\mathcal{M}^n}=\left(\frac{c}{12}\right)^2\{z,w\}\{\bar{z},\bar{w}\}.
\end{eqnarray}

In order to read the entanglement entropy, we only need the information under the limit of $n\to 1$. Expanding $\{z,w\}$ and $\{\bar{z},\bar{w}\}$ near $n=1$, we have
\begin{eqnarray}
\{z,w\}&=&-\frac{2\pi^2}{\beta^2}+(n-1)\frac{4\pi^2\left(1-e^{\frac{2\pi l}{\beta}}\right)^2e^{\frac{4\pi w}{\beta}}}{\beta^2\left(e^{\frac{2\pi w}{\beta}}-e^{\frac{2\pi l}{\beta}}\right)^2\left(e^{\frac{2\pi w}{\beta}}-1\right)^2}+O((n-1)^2),\\
\{\bar{z},\bar{w}\}&=&-\frac{2\pi^2}{\beta^2}+(n-1)\frac{4\pi^2\left(1-e^{\frac{2\pi l}{\beta}}\right)^2e^{\frac{4\pi \bar{w}}{\beta}}}{\beta^2\left(e^{\frac{2\pi \bar{w}}{\beta}}-e^{\frac{2\pi l}{\beta}}\right)^2\left(e^{\frac{2\pi \bar{w}}{\beta}}-1\right)^2}+O((n-1)^2).
\end{eqnarray}
Then we find
\begin{eqnarray}\label{ttn}
\nonumber
\left<T\overline{T}(w,\bar{w})\right>_{\mathcal{M}^n}&=&\left(\frac{c}{12}\right)^2\left[\left(\frac{2\pi^2}{\beta^2}\right)^2+(n-1)\left(-\frac{2\pi^2}{\beta^2}\right)\left(\frac{4\pi^2\left(1-e^{\frac{2\pi l}{\beta}}\right)^2e^{\frac{4\pi w}{\beta}}}{\beta^2\left(e^{\frac{2\pi w}{\beta}}-e^{\frac{2\pi l}{\beta}}\right)^2\left(e^{\frac{2\pi w}{\beta}}-1\right)^2}+h.c.\right)\right]\\
&\;&+O((n-1)^2).
\end{eqnarray}
Plugging (\ref{tt1}) and (\ref{ttn}) into (\ref{deltasn}), then taking the $n\to 1$ limit, we get
\begin{eqnarray}
\delta S(A)=-\mu\left(\frac{c}{12}\right)^2\frac{8\pi^4}{\beta^4}\left(1-e^{\frac{2\pi l}{\beta}}\right)^2\int_{\mathcal{M}}\left[\frac{e^{\frac{4\pi w}{\beta}}}{\left(e^{\frac{2\pi w}{\beta}}-e^{\frac{2\pi l}{\beta}}\right)^2\left(e^{\frac{2\pi w}{\beta}}-1\right)^2}+h.c.\right].
\end{eqnarray}
The integration is a little bit tricky and the details can be found in the appendix \ref{integral}. In the end, we obtain
\begin{eqnarray}
\delta S(A)=\frac{-\mu\pi ^4 c^2 l   \coth \left(\frac{\pi  l}{\beta }\right)}{9 \beta ^3}. \label{mucorrection}
\end{eqnarray}

In the ``low temperature" limit, $\beta\gg l$, the correction to the entanglement entropy (\ref{mucorrection}) is $\frac{-\pi ^3 c^2 \mu }{9 \beta ^2}$.   In the ``high temperature" limit, $\beta\ll l$, the correction is $\frac{-\pi ^4 c^2 l \mu }{9 \beta ^3}$. Actually, at high enough energy, the deformed theory cannot be taken as a local field theory and the above discussion breaks down. Moreover in order to compare with the bulk dual, we have to take the large $c$ limit carefully. It turns out that we should keep $\mu c$ finite in the large $c$ limit\cite{ttatlargec,cutoffads}. Under this limit, the correction of the entanglement entropy is proportional to $c$, which could be compared with the semi-classical action of the gravity.

Recall that the entanglement entropy of $A$ in a CFT with the same setup is
\begin{eqnarray}
S_0(A)=\frac{c}{3}  \log \left(\frac{\beta  }{\pi  \epsilon_0 }\sinh \left(\frac{\pi  l}{\beta }\right)\right),
\end{eqnarray}
with $\epsilon_0$ the CFT cutoff. So to the leading order in $\mu$, we have
\begin{eqnarray}\label{sat}
\nonumber
S(A)&=&S_0(A)+\delta S(A).
\end{eqnarray}

 It is remarkable that although our perturbative computation is to the leading order of $\mu$ and seems work for any temperature, the parameter $\mu$ is of dimension of length square. In the finite temperature case, there is a dimensionless quantity
 \be
 \tmu_\beta=\frac{\mu}{\beta^2},
 \ee
 which cannot be large. In terms of $\tmu_\b$,  the change of the entanglement entropy is
 \be
 \delta S(A)=\frac{-\tmu_\beta \pi^3 c^2}{9}\frac{\pi l}{\beta}\coth\left(\frac{\pi  l}{\beta }\right).
 \ee

 In fact, the leading order correction to the R$\acute{\text{e}}$nyi entropy can also be worked out. The computation of it is more tedious, and the details can be found in the appendix \ref{renyit}. The final result is
 \begin{eqnarray}\label{renyientropyt}
 \nonumber
\delta S_n(A)&=&-\frac{\pi ^4 c^2 l \mu  (n+1) \coth \left(\frac{\pi  l}{\beta }\right)}{18 \beta ^3 n}+\frac{\pi  c^2 \mu  (n-1) (n+1)^2}{576 n^3 \epsilon ^2}\\
\nonumber
&\;&-\frac{\pi ^3 c^2 \mu  (n-1) (n+1)^2 \left(\cosh \left(\frac{2 \pi  l}{\beta }\right)-7\right) \text{csch}^2\left(\frac{\pi  l}{\beta }\right)}{864 \beta ^2 n^3}\\
&\;&+\frac{\pi ^3 c^2 \mu  (n-1) (n+1)^2 \coth ^2\left(\frac{\pi  l}{\beta }\right) \log \left(\frac{\beta  \sinh \left(\frac{\pi  l}{\beta }\right)}{2 \pi  \epsilon }\right)}{36 \beta ^2 n^3}.
\end{eqnarray}
When $n=1$, only the first term survives, and it gives the leading order correction (\ref{mucorrection}) to the entanglement entropy. The second term diverges as $1/\epsilon^2$ and does not depend on $\beta$ and $l$. The third term does not depend on the cutoff $\epsilon$ and can have a finite contribution when $n\neq 1$. The last term has the form $\#\log \big(\frac{\beta  \sinh \left(\frac{\pi  l}{\beta }\right)}{2 \pi  \epsilon }\big)$, recalling that $\log \big(\frac{\beta  \sinh \left(\frac{\pi  l}{\beta }\right)}{2 \pi  \epsilon }\big)$ is the original entanglement entropy.

\subsection{Finite size}\label{finitel}

Another simple case is  a $2D$ deformed CFT at zero temperature but with a finite size $L$. The spatial direction is now compactified, while the time direction is non-compact so the manifold  $\mathcal{M}$  is still an infinitely long cylinder with circumference $L$. We introduce complex coordinate $w=x+i\tau$ and $\bar{w}=x-i\tau$ on the cylinder $\mathcal{M}$, where $\tau\in(-\infty,\infty)$ and $x\in(0,L)$ with the identification $x\sim x+L$. The subsystem $A$ is chosen to be a single interval of length $l<L $ which will be vertical to the axis of the cylinder. The endpoints of $A$ are put at $(w,\bar{w})=(0,0)$ and $(w,\bar{w})=(l,l)$.

The computation procedure is similar to the finite temperature case. Using the map
\begin{eqnarray}
w\to w^\prime=\tan\left(\frac{\pi w}{L}\right),
\end{eqnarray}
which can also map the cylinder to a plane $\mathcal{C}$, we obtain
\begin{eqnarray}\label{tt1l}
\nonumber
\left<T\overline{T}(w,\bar{w})\right>_{\mathcal{M}}&=&\left(\frac{c}{12}\right)^2\{w^\prime,w\}\{\bar{w}^\prime,\bar{w}\}\\
&=&\left(\frac{c}{12}\right)^2\left(\frac{2\pi^2}{L^2}\right)^2.
\end{eqnarray}
The interval $A$ on the cylinder $\mathcal{M}$ is mapped to an interval $A^\prime$ on the plane $\mathcal{C}$ whose endpoints become $(w^\prime,\bar{w}^\prime)=(0,0)$ and $(w^\prime,\bar{w}^\prime)=(\tan \left(\frac{\pi  l}{L}\right),\tan \left(\frac{\pi  l}{L}\right))$. Combining with the map
\begin{eqnarray}
w^\prime\to z=\left(\frac{w^\prime}{w^\prime-\tan \left(\frac{\pi  l}{L}\right)}\right)^{\frac{1}{n}}
\end{eqnarray}
yields a map $w\to z$ which relates $\mathcal{M}^n$ to the plane $\mathcal{C}$. Then we have
\begin{eqnarray}
\left<T\overline{T}(w,\bar{w})\right>_{\mathcal{M}^n}=\left(\frac{c}{12}\right)^2\{z,w\}\{\bar{z},\bar{w}\}.
\end{eqnarray}

Expanding $\{z,w\}$ and $\{\bar{z},\bar{w}\}$ near $n=1$,  we find
\begin{eqnarray}
\nonumber
\{z,w\}&=&\frac{2 \pi ^2}{L^2}+(n-1)\frac{\pi ^2  \sin ^2\left(\frac{\pi  l}{L}\right) \csc ^2\left(\frac{\pi  w}{L}\right) \csc ^2\left(\frac{\pi  (l-w)}{L}\right)}{L^2}+O((n-1)^2),\\
\{\bar{z},\bar{w}\}&=&\frac{2 \pi ^2}{L^2}+(n-1)\frac{\pi ^2  \sin ^2\left(\frac{\pi  l}{L}\right) \csc ^2\left(\frac{\pi  \bar{w}}{L}\right) \csc ^2\left(\frac{\pi  (l-\bar{w})}{L}\right)}{L^2}+O((n-1)^2).
\end{eqnarray}
Then
\begin{eqnarray}\label{ttnl}
\nonumber
\left<T\overline{T}(w,\bar{w})\right>_{\mathcal{M}^n}&=&\left(\frac{c}{12}\right)^2\left[\left(\frac{2 \pi ^2}{L^2}\right)^2+(n-1)\frac{2 \pi ^2}{L^2}\left(\frac{\pi ^2 \sin ^2\left(\frac{\pi  l}{L}\right) \csc ^2\left(\frac{\pi  w}{L}\right) \csc ^2\left(\frac{\pi  (l-w)}{L}\right)}{L^2}+h.c.\right)\right]\\
&\;& +O((n-1)^2).
\end{eqnarray}

Plugging (\ref{tt1l}) and (\ref{ttnl}) into (\ref{deltasn}), then taking the $n\to 1$ limit, we get
\begin{eqnarray}
\delta S(A)=\frac{\mu \pi ^4 c^2   \sin ^2\left(\frac{\pi  l}{L}\right)}{72 L^4}\int_{\mathcal{M}}\left[\csc ^2\left(\frac{\pi  w}{L}\right) \csc ^2\left(\frac{\pi  (l-w)}{L}\right)+h.c.\right].
\end{eqnarray}
The integral involved is
\begin{eqnarray}
\int_{\mathcal{M}}\csc ^2\left(\frac{\pi  w}{L}\right) \csc ^2\left(\frac{\pi  (l-w)}{L}\right)=\int_{-\infty}^{\infty}d\tau \int_0^Ldx\csc ^2\left(\frac{\pi  (x+i\tau))}{L}\right) \csc ^2\left(\frac{\pi  (l-(x+i\tau)))}{L}\right).\;\;\;\;
\end{eqnarray}
We first do the $x$ integral. Fortunately the primitive function can be found, which is
\begin{eqnarray}
\nonumber
&\;&-\frac{8 i L e^{\frac{2 i \pi  l}{L}} \left(1+e^{\frac{2 i \pi  l}{L}}\right) \left(\log \left(1-e^{\frac{2 i \pi  (x+i \tau )}{L}}\right)-\log \left(1-e^{\frac{2 i \pi  (-l+i \tau +x)}{L}}\right)\right)}{\pi  \left(-1+e^{\frac{2 i \pi  l}{L}}\right)^3}\\
&\;&+\frac{8 i L e^{\frac{2 \pi  (\tau +i l)}{L}} \left(2 e^{\frac{2 \pi  (\tau +i l)}{L}}-e^{\frac{2 i \pi  (l+x)}{L}}-e^{\frac{2 i \pi  x}{L}}\right)}{\pi  \left(-1+e^{\frac{2 i \pi  l}{L}}\right)^2 \left(-e^{\frac{2 \pi  \tau }{L}}+e^{\frac{2 i \pi  x}{L}}\right) \left(e^{\frac{2 \pi  (\tau +i l)}{L}}-e^{\frac{2 i \pi  x}{L}}\right)}.
\end{eqnarray}
The term on the second line has no contribution since plugging $x=0$ or $x=L$ into it gives the same result. The term on the first line  has no contribution as well since the two $\log$ terms always cancel each other. This is very different from the finite temperature case. Thus we learn that
\begin{eqnarray}
\int_{\mathcal{M}}\csc ^2\left(\frac{\pi  w}{L}\right) \csc ^2\left(\frac{\pi  (l-w)}{L}\right)=0,
\end{eqnarray}
which means
\begin{eqnarray}
\delta S(A)=0.
\end{eqnarray}
So to the leading order of $\mu$, the entanglement entropy of $A$ is still
\begin{eqnarray}\label{sal}
S(A)=\frac{c}{3}  \log \left(\frac{L }{\pi  \epsilon_0 }\sin \left(\frac{\pi  l}{L}\right)\right),
\end{eqnarray}
with $\epsilon_0$ the CFT cutoff.

On the contrary, the leading order correction to the R$\acute{\text{e}}$nyi entropy in this case is not vanishing. The computation is similar to the finite-temperature case, and the details  can be found in the appendix \ref{renyil}. And the result is
\begin{eqnarray}
\delta S_n(A)&=&\frac{\pi  c^2 \mu  (n-1) (n+1)^2}{576 n^3 \epsilon ^2}\nn\\
&\;&-\frac{\pi ^3 c^2 \mu  (n-1) (n+1)^2 \left(11 \cos \left(\frac{2 \pi  l}{L}\right)+19\right) \csc ^2\left(\frac{\pi  l}{L}\right)}{864 L^2 n^3}\nn\\
&\;&+\frac{\pi ^3 c^2 \mu  (n-1) (n+1)^2 \cot ^2\left(\frac{\pi  l}{L}\right) \log \left(\frac{L \sin \left(\frac{\pi  l}{L}\right)}{2 \pi  \epsilon }\right)}{36 L^2 n^3}.
\end{eqnarray}
When $n=1$, it is vanishing as we expect. Let us compare it with (\ref{renyientropyt}):  the quadratic divergent terms ($1/\epsilon^2$) are the same, which is independent of the finite temperature or finite size; their logarithmic terms are the same under the identification $L \leftrightarrow i\b$. The main difference between them is that $\delta S_n(A)$ in the finite $T$ case has an additional term
\begin{eqnarray}
-\frac{\pi ^4 c^2 l \mu  (n+1) \coth \left(\frac{\pi  l}{\beta }\right)}{18 \beta ^3 n},
\end{eqnarray}
which is nonzero when $n=1$.

\section{Gravity dual}

The AdS/CFT correspondence\cite{Maldacena:1997re} states that the gravitational theory living in the bulk is dual to a CFT living on the asymptotic boundary of the AdS spacetime. Especially the Ryu-Takayanagi formula \cite{rtformula,Ryu:2006ef} relates the entanglement entropy in the CFT with the area of the corresponding minimal surface in the gravitational theory. The holographic entanglement entropy could be understood as a generalized gravitational entropy\cite{proofhee}.

For the AdS$_3$/CFT$_2$ correspondence, it has been found that after imposing appropriate asymptotic boundary condition\cite{Brown:1986nw}, the AdS$_3$ gravity could be dual to a 2D CFT with central charge\cite{Strominger:1997eq}
\be
c=\frac{3R}{2G}.
\ee
The authours of \cite{verlinde} proposed that under $T\overline{T}$ deformation the bulk dual gravitational theory should be defined  by moving the asymptotic boundary inwards with the radius being at
\be
r_c^2=\frac{6R^4}{\mu \pi c}.
\ee
Here $R$ is the AdS radius and $c$ is the central charge of the dual CFT. In this case, we expect that the holographic entanglement entropy is still given by the RT-formula,
\begin{eqnarray}\label{rtformu}
S(A)=\frac{\text{Area}\;\text{of}\;\gamma_A}{4G},
\end{eqnarray}
where $\gamma_A$ is the minimal surface in the bulk whose boundary is given by $\partial A$. In the cases we are considering $A$ is an interval, and $\gamma_A$ is the geodesic whose endpoints coincide with $A$'s.

\subsection{BTZ black hole}\label{btzblackhole}
A $2d$ CFT at high temperature is dual to a BTZ black hole. According to \cite{verlinde} the $T\overline{T}$ deformed CFT at high temperature is naturally dual to a BTZ black hole with a radial cutoff. The metric of Euclidean BTZ black hole is
\begin{eqnarray}\label{btz}
ds^2=\frac{r^2-r_+^2}{R^2}dt^2+\frac{R^2}{r^2-r_+^2}dr^2+r^2dx^2,
\end{eqnarray}
with $R$ the AdS radius, $r_+$ the position of the horizon, $t$ compactified as $t\sim t+\beta$. $\beta=\frac{2\pi R^2}{r_+}$ is the temperature of the black hole and the corresponding CFT.

Originally the boundary is located at $r\to \infty$. Now we move the boundary inwards to $r=r_c$ where the $T\overline{T}$ deformed CFT lives. On this new boundary, the metric is
\begin{eqnarray}
\nonumber
ds^2_b&=&\frac{r_c^2-r_+^2}{R^2}dt^2+r_c^2dx^2\\
&=&\frac{r_c^2-r_+^2}{R^2}\left(dt^2+\frac{R^2r_c^2}{r_c^2-r_+^2}dx^2\right).
\end{eqnarray}
The black hole temperature is still $\beta$, which means the temperature of the $T\overline{T}$ deformed CFT is also $\beta$. So $t$ is the physical time of the deformed CFT, whose physical metric shall be
\begin{eqnarray}\label{pmetric}
ds_p^2=dt^2+\frac{R^2r_c^2}{r_c^2-r_+^2}dx^2.
\end{eqnarray}

At some time $t_0$, we put the endpoints of the subsystem $A$ at $(t,x)=(t_0,0)$ and $(t,x)=(t_0,\delta x)$. According to (\ref{pmetric}), the length of $A$ is
\begin{eqnarray}\label{lengthl}
l=\frac{\delta xRr_c}{\sqrt{r_c^2-r_+^2}}.
\end{eqnarray}
What we are left to do is to find the geodesic distance $\lambda$ between $(r,t,x)=(r_c,t_0,0)$ and $(r,t,x)=(r_c,t_0,\delta x)$. To achieve this we define the new coordinates
\begin{eqnarray}
r=r_+\cosh \rho,\;\;t=\frac{ R^2\theta}{r_+},\;\;x=\frac{R \tau}{r_+},
\end{eqnarray}
following which the metric (\ref{btz}) becomes
\begin{eqnarray}
ds^2=R^2\left(\sinh^2 \rho d\theta^2+d\rho^2+\cosh^2 \rho d\tau^2\right),
\end{eqnarray}
which is the Euclidean version of global $AdS_3$ metric. The endpoints become $(\rho,\theta,\tau)=(\rho_c,\theta_0,0)$ and $(\rho,\theta,\tau)=(\rho_c,\theta_0,\frac{ r_+\delta x}{R})$ with
\begin{eqnarray}\label{rcandrho}
r_c=r_+\cosh \rho_c.
\end{eqnarray}
Now the geodesic distance $\lambda$ can be easily found:
\begin{eqnarray}
\cosh\left(\frac{\lambda}{R}\right)=1+2\cosh^2 \rho_c \sinh^2 \frac{r_+\delta x}{2R}.
\end{eqnarray}
Plugging (\ref{rcandrho}) and (\ref{lengthl}) into it, we find
\begin{eqnarray}
\cosh\left(\frac{\lambda}{R}\right)=1+2\left(\frac{r_c}{r_+}\right)^2\sinh^2\left(\frac{\pi l}{\beta}\sqrt{1-\left(\frac{r_+}{r_c}\right)^2}\right).
\end{eqnarray}
When $r_c\gg r_+$, we can expand $\lambda$ in terms of $r_+/r_c$ and obtain
\begin{eqnarray}\label{lambdat}
\frac{\lambda}{4G}=\frac{R}{2G}\log\left(\frac{\beta r_c\sinh\left(\frac{\pi l}{\beta}\right)}{\pi R^2}+\frac{\pi R^2}{\beta r_c\sinh\left(\frac{\pi l}{\beta}\right)}-\frac{2\pi^2 R^2 l \cosh\left(\frac{\pi l}{\beta}\right)}{\beta^2 r_c}+O\left(\left(\frac{r_+}{r_c}\right)^2\right)\right),
\end{eqnarray}
where we have used $\beta=\frac{2\pi R^2}{r_+}$ to replace $r_+$ by $\frac{2\pi R^2}{\beta}$. If we consider the ``high temperature" case $\beta <l$, the second term in the parenthesis can be naturally ignored since it is much smaller than the third term. On the other hand since $r_c$ is very large, we can treat the third term as a small quantity compared with the first term. This leads to
\bea
\frac{\lambda}{4G}&=&\frac{R}{2G}\log\left(\frac{\beta r_c\sinh\left(\frac{\pi l}{\beta}\right)}{\pi R^2}\right)-\frac{R}{2G}\frac{2\pi^3R^4l\coth\left(\frac{\pi l}{\beta}\right)}{\beta^3 r^2_c}\nn\\
&=&\frac{c}{3}\log\left(\frac{\beta r_c\sinh\left(\frac{\pi l}{\beta}\right)}{\pi R^2}\right)-\frac{\pi^4 c^2 \mu l}{9\beta^3}\coth\left(\frac{\pi l}{\beta}\right)
\eea
after considering the relations $c=3R/2G$ and (\ref{muandrc}). Now as the cutoff boundary is at $r_c$ so the corresponding cutoff in the field theory is $\epsilon=R^2/r_c$, then we find the perfect match with the field theory result.


\subsection{Global AdS}
A $2d$ CFT at zero temperature with the spatial direction compactified lives on the asympotic boundary of the global $AdS_3$. So the $T\overline{T}$ deformed CFT which we considered in section (\ref{finitel}) is dual to the global $AdS_3$ with a radial cutoff. The metric of global $AdS_3$ is
\begin{eqnarray}
ds^2=R^2\left(-\cosh^2\rho dt^2+d\rho^2+\sinh^2 \rho d\phi^2\right),
\end{eqnarray}
with $\phi$ compactified as $\phi\sim \phi+2\pi$. We put the boundary at $\rho=\rho_c$, on which the $T\overline{T}$ deformed CFT lives. At some time $t_0$, the endpoints of the subsystem $A$ are put at $(t,\phi)=(t_0,0)$ and $(t,\phi)=(t_0, \delta \phi)$. Suppose that the total length of the quantum system is $L$, then the length of $A$ is given by
\begin{eqnarray}\label{landphi}
l=\frac{\delta \phi L}{2\pi}.
\end{eqnarray}

The geodesic distance $\lambda$ between $(\rho,t,\phi)=(\rho_c,t_0,0)$ and $(\rho,t,\phi)=(\rho_c,t_0,\delta \phi)$ is given by
\begin{eqnarray}
\cosh \left(\frac{\lambda }{R}\right)=1+2\sinh^2\rho_c \sin^2\left(\frac{\pi l}{L}\right),
\end{eqnarray}
where we have used (\ref{landphi}) to replace $\delta \phi$ by $2\pi l/L$.
When $\rho_c\gg1\leftrightarrow \sinh \rho_c\gg1$, we can expand $\lambda$ in terms of $1/ \sinh \rho_c$ and obtain
\begin{eqnarray}\label{lambdal}
\frac{\lambda}{4G}=\frac{R}{2G}\log\left(2\sinh \rho_c \sin\left(\frac{\pi l}{L}\right)+\frac{1}{2\sinh \rho_c \sin\left(\frac{\pi l}{L}\right)}+O\left(\left(\frac{1}{\sinh\rho_c}\right)^2\right)\right).
\end{eqnarray}
Now  there is no other correction except that the cutoff surface is moved inward. This fact is in accordance with the fact that there is no correction to the entanglement entropy from the $T\overline{T}$ deformation in the finite size CFT in the leading order of $\mu$.

After the careful calculations on the bulk side, we notice that the main difference between these two cases lies on the difference between (\ref{lengthl}) and (\ref{landphi}). (\ref{lengthl}) says that $\delta x$ depends on the cutoff $r_c$ when $r_+\neq0$ (i.e. $1/\beta\neq0$), while (\ref{landphi}) shows that $\delta \phi$ does not depend on the cutoff. In the finite temperature case the leading order correction comes actually from the $r_c$ dependence of $\delta x$. 

\section{More general holographic picture}

In the above discussion on the holographic entanglement entropy, we actually assumed the RT prescription. This expectation turns out to be good. However, for the single-interval R\'enyi entropy, we need to consider the backreaction of the twist operator\cite{proofhee, Dong:2016fnf}. In the following, we try to argue that the holographic picture is still true for general configurations, using the method developed in \cite{Skenderis:1999nb, Hung:2011nu, Faulkner:2013yia, Barrella:2013wja}.

We start from the $T\overline{T}$-deformed CFT defined on the boundary metric
\begin{equation}
ds^2=g_{ab}dx^adx^b.
\end{equation}
It is dual to the  gravitational theory living on a compact sub-region of AdS. The metric of the bulk configuration could be \begin{equation}
ds^2=\frac{dr^2}{r^2}+r^2g_{ab}dx^adx^b.
\end{equation}
We have set $R_{AdS}=1$. The Poincare coordinate is recovered by setting $\xi=1/r$.  In  the Fefferman-Graham gauge, the metric is expanded as
\begin{equation}
ds^2=\frac{d\rho^2}{4\rho^2}+\frac{g_{ab}}{\rho}dx^adx^b.
\end{equation}
Having fixed the leading order $g_{(0)}$, the metric above is characterized by the stress tensor of the classical Liouville field\cite{Krasnov:2000zq}. In other words,  the classical gravitational solution is characterized by the stress tensor, which is determined by the conformal weights and  the accessory parameters in particular. In general, it is hard to find the explicit form of the metric. In the following discussion we denote the $x^a$ in Poincare coordinate as $z,\bar{z}$, while the $x^a$ in FG coordinate is denoted as $w, \bar{w}$.

For the $T\overline{T}$-deformed holographic CFT, there is a certain regulator surface at a fixed radial position.  We choose the regulator surface in the Poincare coordinate, so that the induced metric of the surface coincides with the one in CFT. The regulator surface is located at
\begin{equation}
\xi_c\approx be^{\phi},
\end{equation}
where $\phi$ is the classical Liouville field, relating to the Weyl factor, and
\begin{equation}
b^2=\frac{\mu c}{24\pi}.
\end{equation}
There is a coordinate transformation between the FG coordinate and the Poincare coordinate\cite{Krasnov:2001cu}
\begin{equation}
\xi=\frac{\rho^{1/2}e^{-\phi}}{1+\rho e^{-2\phi}a^2},
\end{equation}
where $a=\partial \phi$.

The semi-classical action of the gravitational theory is
\begin{equation}
I=I_{EH}+I_{GH}+I_{CT},
\end{equation}
including the Einstein-Hilbert term plus a negative cosmological constant, the Gibbons-Hawking term and the counter term.
The counter term cancels the power-law divergence in the bulk integral and the boundary integral. More concretely, the on-shell Einstein-Hilbert action reduces to
\begin{equation}
I_{EH}=-\frac{c}{96\pi}\int dzd\bar{z}\xi_c^{-2},
\end{equation}
\begin{equation}
\xi_c^{-2}=a^4 b^2 e^{-2 \phi }+2
   a^2-\frac{e^{2 \phi }}{b^2}.
\end{equation}
The Gibbons-Hawking term and the counter term give
\begin{equation}
I_{GH}+I_{CT}=-\frac{c}{96\pi}\int dzd\bar{z}(\frac{e^{2 \phi }}{b^2}+8\partial\bar{\partial}\phi).
\end{equation}
The final on-shell action is
\begin{equation}
I=-\frac{c}{96\pi}\int dzd\bar{z}(2a^2+a^4 b^2 e^{-2 \phi }-8\partial\bar{\partial}\phi).
\end{equation}
Note that as $b\rightarrow 0$, the action above is just the Liouville action, and the changes in the choice of cut-off surface is sub-leading in $b$. Note also that there is an ambiguity in the choice of the counter term, so the linear order change in the bulk action does matter, leaving the other potential terms depending on certain regularization prescription.

To go further, we turn to calculate the above action in the FG gauge with a proper regulator  surface. It turns out that the last part vanishes and the first part becomes
\begin{equation}
I_{1}=-n\frac{c}{96\pi}\int dwd\bar{w} 4\sqrt{T_L\bar{T}_L}.
\end{equation}
The result above can be understood as follows
\begin{equation}
-\frac{c}{96\pi}\int dzd\bar{z} 2a_z^2=-\frac{c}{96\pi}\int dwd\bar{w}e^{-2\phi}e^{2\phi} 2a_w^2=-n\frac{c}{96\pi}\int dwd\bar{w} 4\sqrt{ T_L\bar{T}_L}.
\end{equation}
The $T_L $ is the Liouville stress tensor, related to the vacuum expectation value of the CFT stress tensor by
\begin{equation}
\bra T \ket_{CFT}=-\frac{1}{2l_p}T_L.
\end{equation}
The action above can be used to get the HRE when certain conformal transformation has been made\cite{Hung:2011nu}, and $l_p=8\pi G$.

The remaining part, which is associated to the regulator surface and gives the correction to the HRE, can be calculated by
\begin{equation}
I_{2}=-\frac{c}{96\pi}\int dzd\bar{z}a^4 b^2 e^{-2 \phi }=-\frac{cb^2}{96\pi}\int dwd\bar{w} a^4_w e^{4\phi} e^{-2\phi}e^{-2\phi}=-n\frac{cb^2}{96\pi}\int dwd\bar{w}4 T_L\bar{T}_L.
\end{equation}
Considering the fact that
\begin{equation}
c=\frac{12\pi}{l_p}=\frac{3}{2G},
\end{equation}
we find that the integrand from field theory side is
\begin{equation}
-\frac{cb^2}{96\pi}4T_L\bar{T}_L=-\frac{\mu c^2}{576\pi^2}T_L\bar{T}_L.
\end{equation}
Recall the involved partition function $Z_n$ from the conformal perturbation theory, where the linear term in $\mu$ is just
\begin{equation}
-\mu\bra T\bar{T} \ket_{CFT} =-\mu\bra T\ket_{CFT} \bra \bar{T} \ket_{CFT}=-\frac{\mu}{4l_p^2}T_L  \bar{T}_L =-\frac{\mu c^2}{576\pi^2}T_L\bar{T}_L.
\end{equation}
Thus at the linear level, the QFT partition function calculated by the conformal perturbation theory matches with the gravitational result.

Note that the discussion may apply to the more general cases than the single-interval R\'enyi entropy. For example, for the two-interval case\cite{Barrella:2013wja,Chen:2013kpa} and the single-interval on a torus case\cite{Barrella:2013wja,Chen:2014unl}, the leading order correction in $\mu$ to the R\'enyi entropy should match with the holographic computation as well.


\section{Discussion}

In this papar we have calculated the entanglement entropy of a single interval on a cylinder in the $T\overline{T}$-deformed CFT. We find that the leading order correction to the entanglement entropy is nonzero in the finite temperature case while it is vanishing in the finite size case. In the dual bulk side it is expected naively that moving inwards will certainly change the geodesic distances which means the leading order correction should be nonzero in both cases. However in the finite size case, the change of the boundary could actually be taken into account by a different cutoff. On the contrary, in the high temperature case, such a change do modify the geodesic distance. Our study supports the conjecture proposed in \cite{verlinde}.

Unlike the work done in \cite{eeandtt}, our field theory results are only valid when $\mu\to 0$. To obtain the finite $\mu$ results, we need to know the partition function of the theory $\mathcal{T}^{(\mu)}$ on $\mathcal{M}$ and $\mathcal{M}^n$, which is a much harder job. It would be definitely interesting to study this issue. 
On the gravity side, the discussion in the present work relies also  on  the condition that $\mu$ is very small. In the finite $\mu$ case, it is not clear if the RT prescription can be applied naively. To determine whether the RT formula is still valid or not, one should go to the non-perturbative level. It is interesting to consider the full version of this duality, saying arbitrary geometry and finite deformation. 

The holographic entanglement entropy in the standard $AdS/CFT$ correspondence has brought us many new understandings of the holographic duality. The new duality proposed in \cite{verlinde} is fascinating, and provides a new window to study various problems in the $AdS/CFT$ holography , like holographic entanglement entropy, bulk reconstruction, holographic complexity etc.. We wish to address these issues in the future.


\section*{Acknowledgments}

We are grateful to Han Liu for the participation at the early stage of the project. We would like to thank Yan-jun Liu for valuable discussions.  The work is supported in part by NSFC Grant No. 11275010, No. 11325522, No.~11335012
and No. 11735001.

\appendix
\renewcommand{\appendixname}{Appendix~\Alph{section}}
\section{The Integral}\label{integral}
\begin{figure}
\centering
\subfigure[$x>0$]{\includegraphics[width=2.3in]{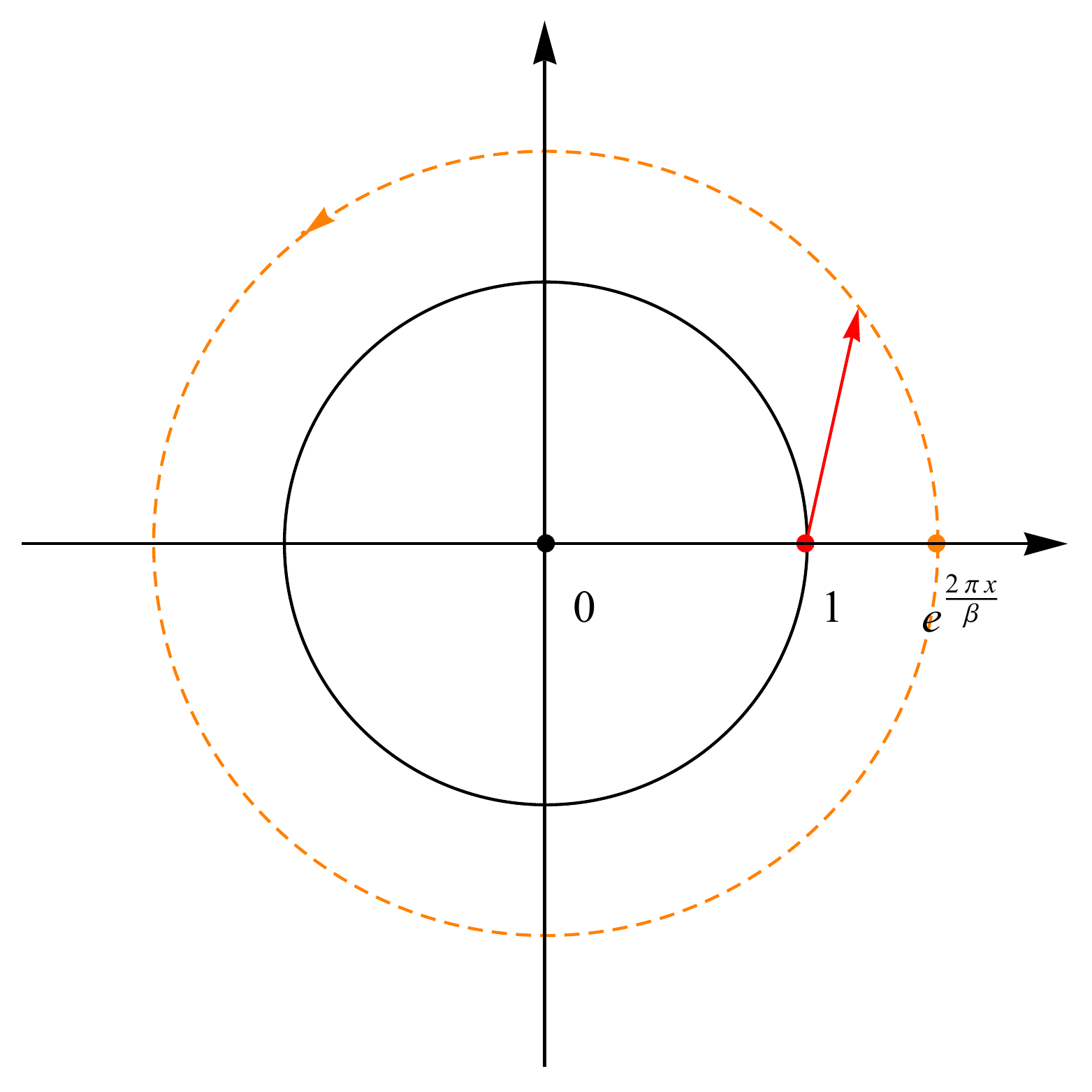}}
\subfigure[$x<0$]{\includegraphics[width=2.3in]{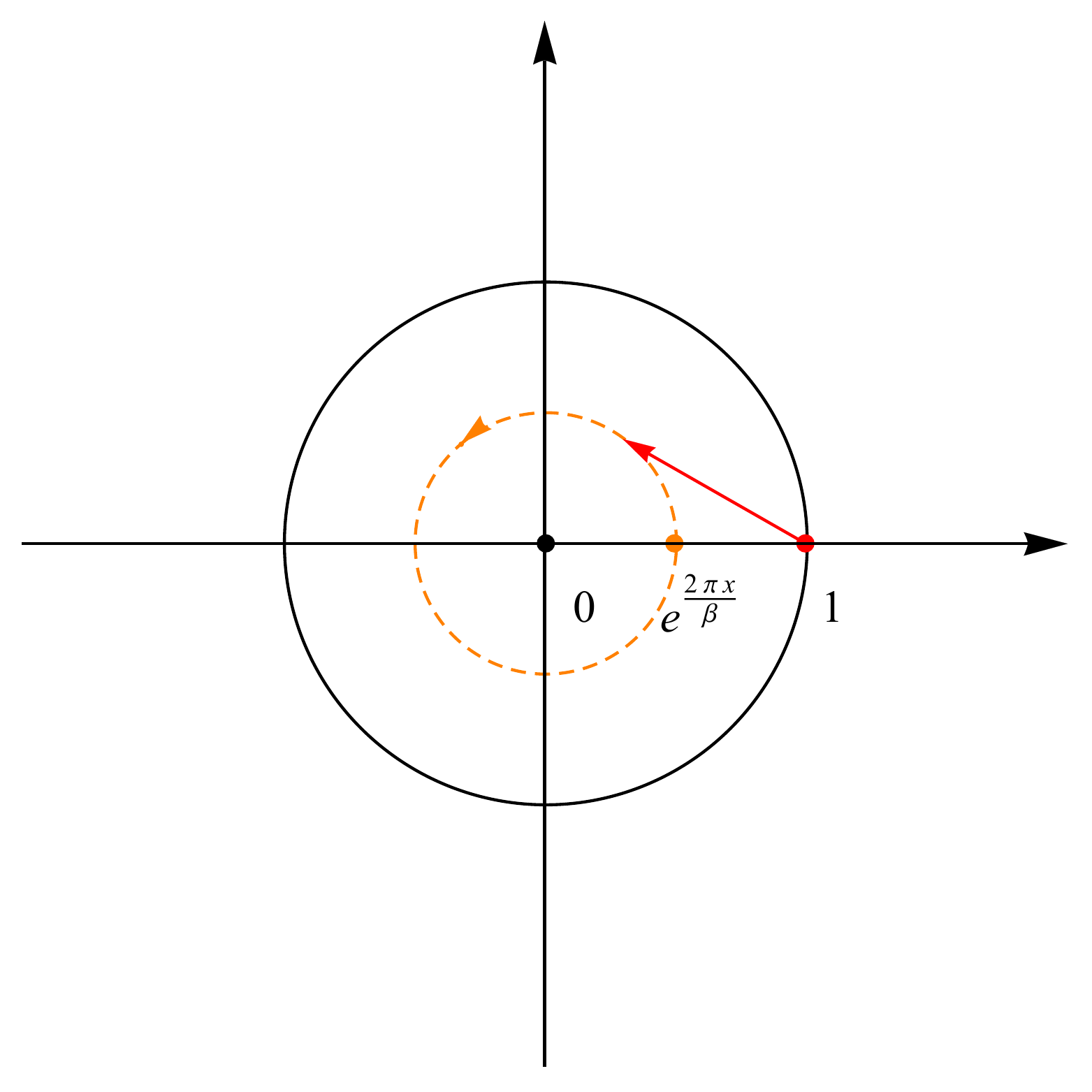}}
\captionsetup{font={small}}
\caption{$e^{\frac{2 \pi  (x+i \tau )}{\beta }}$ runs on the orange circle clockwise, and $(e^{\frac{2 \pi  (x+i \tau )}{\beta }}-1)$ is represented by the red arrow. When moving the head of the arrow around the orange circle once: in (a) the argument of the red arrow is added by $2\pi$, so $\log \left(e^{\frac{2 \pi  (x+i \tau )}{\beta }}-1\right)\Big|_{\tau=0}^{\tau=\beta}=2\pi i$; in (b) the argument of the red arrow doesn't change, so $\log \left(e^{\frac{2 \pi  (x+i \tau )}{\beta }}-1\right)\Big|_{\tau=0}^{\tau=\beta}=0$.}
\label{fig1}
\end{figure}

In this appendix, we present the details of the integration in section (\ref{finitet}). In order to work out the integral
\begin{eqnarray}
\int_{\mathcal{M}}\frac{e^{\frac{4\pi w}{\beta}}}{\left(e^{\frac{2\pi w}{\beta}}-e^{\frac{2\pi l}{\beta}}\right)^2\left(e^{\frac{2\pi w}{\beta}}-1\right)^2}=\int_{-\infty}^{\infty} dx\int_0^\beta d\tau\frac{e^{\frac{4\pi (x+i\tau)}{\beta}}}{\left(e^{\frac{2\pi (x+i\tau)}{\beta}}-e^{\frac{2\pi l}{\beta}}\right)^2\left(e^{\frac{2\pi (x+i\tau)}{\beta}}-1\right)^2},
\end{eqnarray}
we first do the $\tau$ integral. Luckily the primitive function can be found to be
\begin{eqnarray}
\nonumber
\frac{i \beta  \left(\frac{e^{\frac{2 \pi  l}{\beta }}-1}{-1+e^{\frac{2 \pi  (x+i \tau )}{\beta }}}+\frac{e^{\frac{2 \pi  l}{\beta }}-e^{\frac{4 \pi  l}{\beta }}}{e^{\frac{2 \pi  l}{\beta }}-e^{\frac{2 \pi  (x+i \tau )}{\beta }}}-\left(e^{\frac{2 \pi  l}{\beta }}+1\right) \left(\log \left(e^{\frac{2 \pi  (x+i \tau )}{\beta }}-1\right)-\log \left(e^{\frac{2 \pi  (x+i \tau )}{\beta }}-e^{\frac{2 \pi  l}{\beta }}\right)\right)\right)}{2 \pi  \left(e^{\frac{2 \pi  l}{\beta }}-1\right)^3}.\;\;
\end{eqnarray}
Plugging $\tau=0$ or $\tau=\beta$ into the term
\begin{eqnarray}
\frac{e^{\frac{2 \pi  l}{\beta }}-1}{-1+e^{\frac{2 \pi  (x+i \tau )}{\beta }}}+\frac{e^{\frac{2 \pi  l}{\beta }}-e^{\frac{4 \pi  l}{\beta }}}{e^{\frac{2 \pi  l}{\beta }}-e^{\frac{2 \pi  (x+i \tau )}{\beta }}}
\end{eqnarray}
gives the same result, so this term has no contribution. The term we shall analyze carefully is
\begin{eqnarray}
-\left(e^{\frac{2 \pi  l}{\beta }}+1\right) \left(\log \left(e^{\frac{2 \pi  (x+i \tau )}{\beta }}-1\right)-\log \left(e^{\frac{2 \pi  (x+i \tau )}{\beta }}-e^{\frac{2 \pi  l}{\beta }}\right)\right).
\end{eqnarray}
Let us first focus on
\begin{eqnarray}
\log \left(e^{\frac{2 \pi  (x+i \tau )}{\beta }}-1\right).
\end{eqnarray}
Fixing $x$, when $\tau$ runs from $0$ to $\beta$, $e^{\frac{2 \pi  (x+i \tau )}{\beta }}$ runs around the origin once with circular orbit of radius $e^{\frac{2 \pi  x}{\beta }}$. If the radius $e^{\frac{2 \pi  x}{\beta }}>1$, $e^{\frac{2 \pi  (x+i \tau )}{\beta }}$ will run around $1$ once, which means that $\log \left(e^{\frac{2 \pi  (x+i \tau )}{\beta }}-1\right)$ will contribute $2\pi i$. It is demonstrated explicitly in Fig.\ref{fig1}. So we have
\begin{eqnarray}
\nonumber
\log \left(e^{\frac{2 \pi  (x+i \tau )}{\beta }}-1\right)\Big|_{\tau=0}^{\tau=\beta}&=&\left\{
\begin{aligned}
0, \;\;\;\;x<0 \leftrightarrow e^{\frac{2 \pi  x}{\beta }}<1\\
2\pi i, \;\;\;\;x>0 \leftrightarrow e^{\frac{2 \pi  x}{\beta }}>1,
\end{aligned}
\right.\\
\end{eqnarray}

\begin{figure}
\centering
\subfigure[$x<0$]{\includegraphics[width=2.12in]{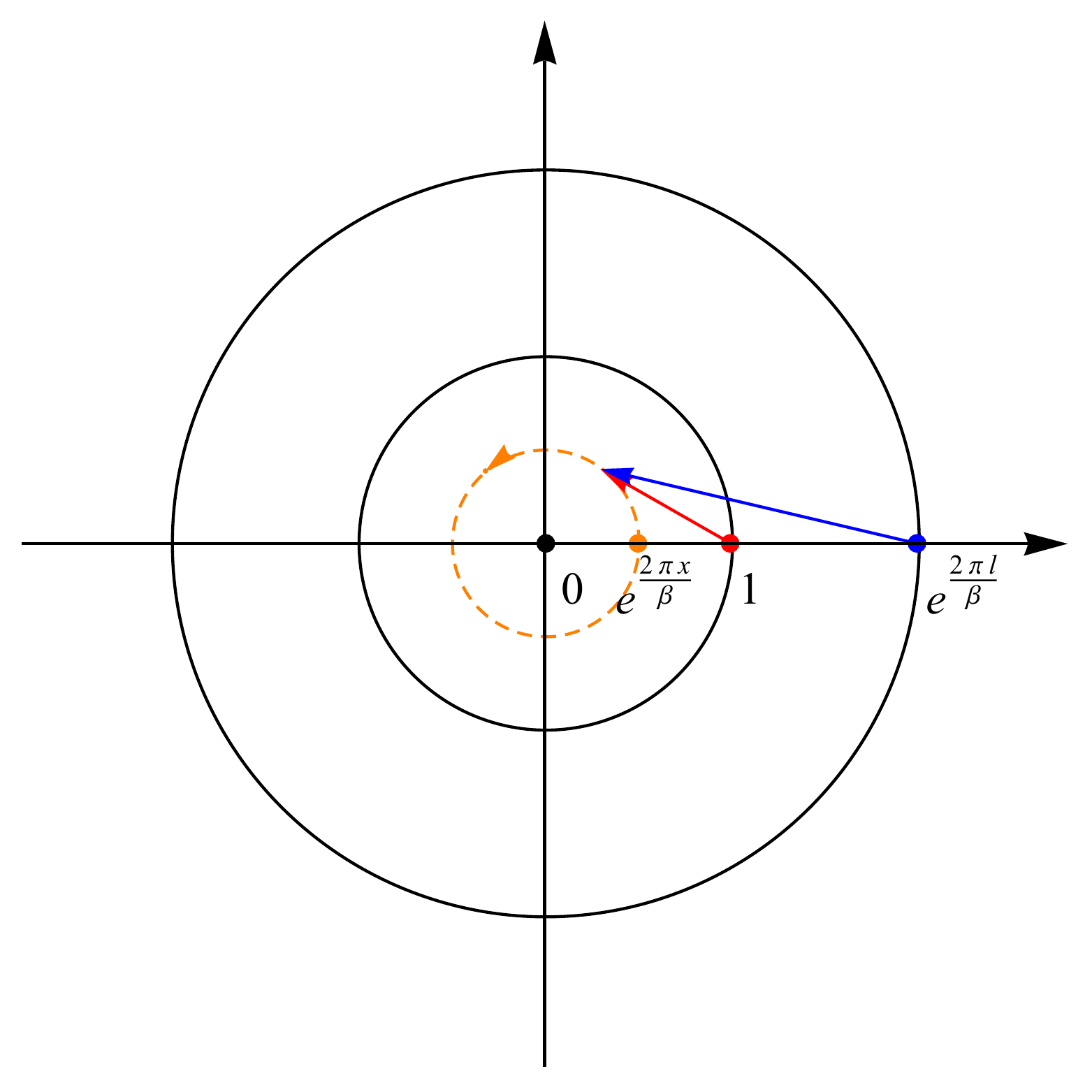}}
\subfigure[$0<x<l$]{\includegraphics[width=2.12in]{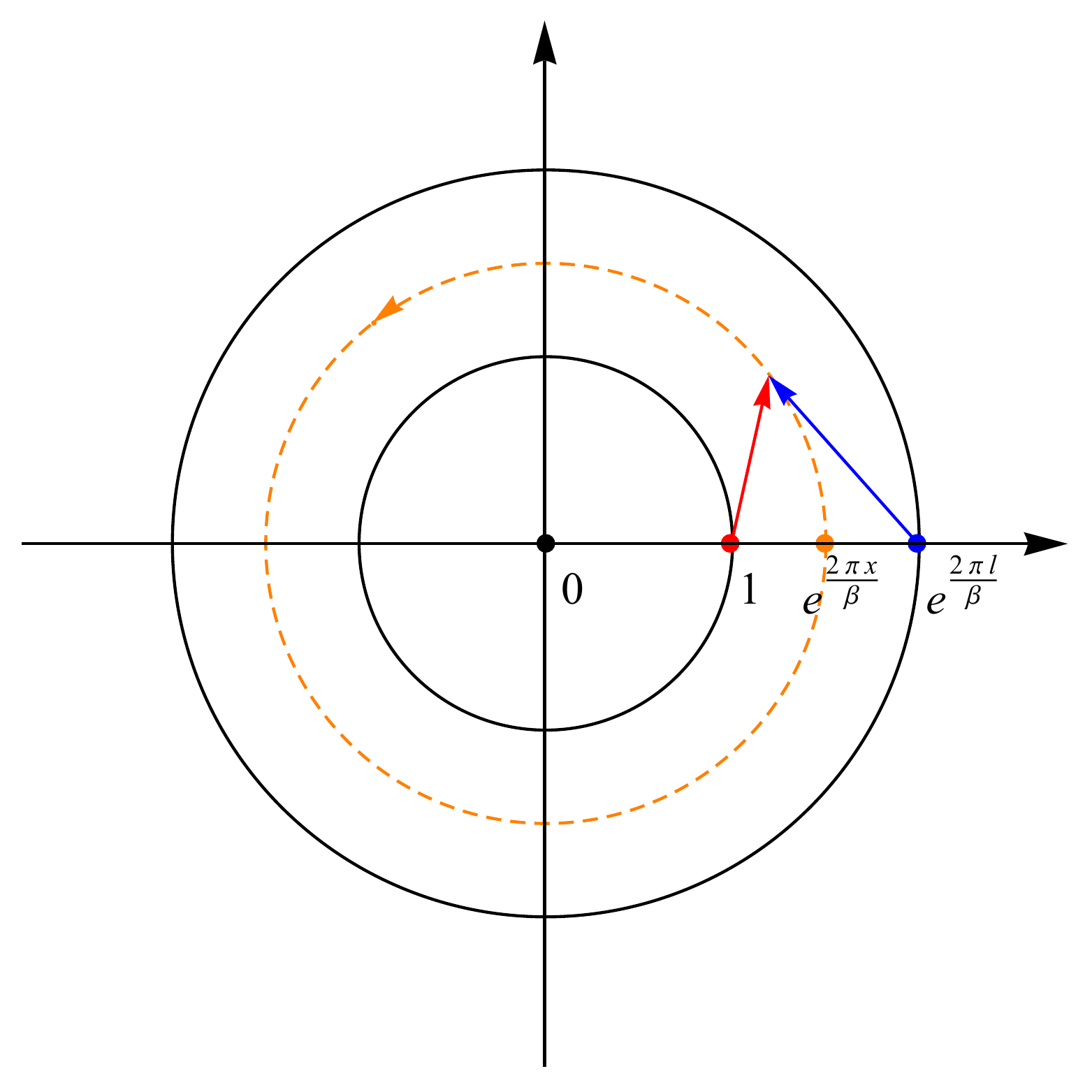}}
\subfigure[$l<x$]{\includegraphics[width=2.12in]{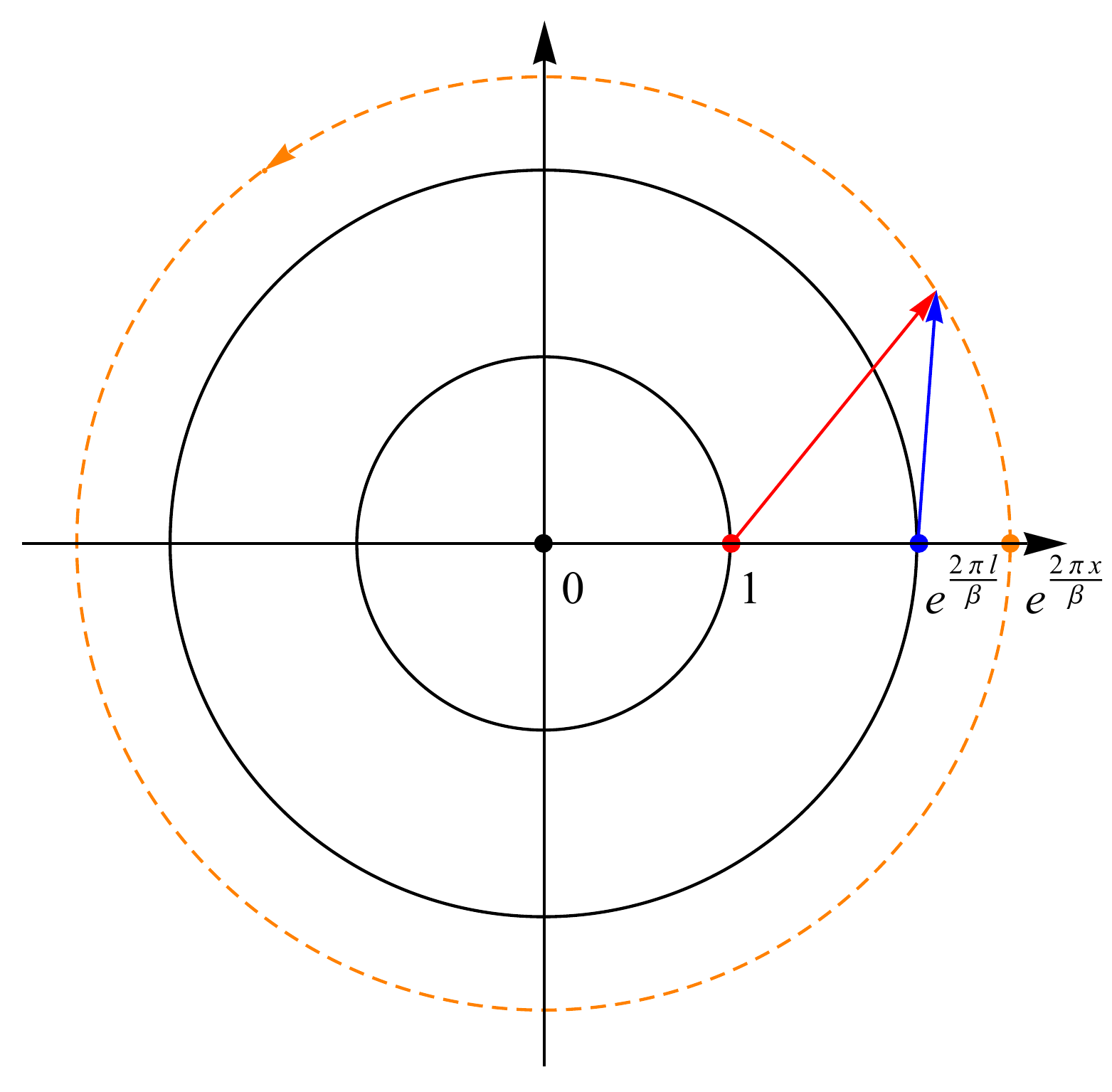}}
\captionsetup{font={small}}
\caption{Again $e^{\frac{2 \pi  (x+i \tau )}{\beta }}$ runs on the orange circle clockwise. $(e^{\frac{2 \pi  (x+i \tau )}{\beta }}-1)$ is represented by the red arrow, and $(e^{\frac{2 \pi  (x+i \tau )}{\beta }}-e^{\frac{2 \pi  l}{\beta }})$ is represented by the blue arrow. When moving the heads of the arrows around the orange circle once: in (a) the arguments of both arrows won't change, which means $g(x)=0$; in (b) the argument of the red arrow is added by $2\pi$, while the argument of the bule arrow doesn't change, so we have $g(x)=2\pi i$; in (c) the arguments of both arrows are added by $2\pi$, but their contributions cancle each other, so $g(x)=0$.}
\label{fig2}
\end{figure}

For the other logarithmic function,  the  discussion is similar:
\begin{eqnarray}
\nonumber
 \log \left(e^{\frac{2 \pi  (x+i \tau )}{\beta }}-e^{\frac{2 \pi  l}{\beta }}\right)\Big|_{\tau=0}^{\tau=\beta}&=&\left\{
\begin{aligned}
0, \;\;\;\;x<l \leftrightarrow e^{\frac{2 \pi  x}{\beta }}<e^{\frac{2 \pi  l}{\beta }}\\
2\pi i, \;\;\;\;x>l \leftrightarrow e^{\frac{2 \pi  x}{\beta }}>e^{\frac{2 \pi  l}{\beta }}.
\end{aligned}
\right.
\end{eqnarray}
Consequently, we have
\begin{eqnarray}\label{logterm}
g(x)\equiv\log \left(e^{\frac{2 \pi  (x+i \tau )}{\beta }}-1\right)-\log \left(e^{\frac{2 \pi  (x+i \tau )}{\beta }}-e^{\frac{2 \pi  l}{\beta }}\right)\Big|_{\tau=0}^{\tau=\beta}&=&\left\{
\begin{aligned}
0&,& \;\;\;\;x<0 \\
2\pi i&,& \;\;\;\;0<x<l\\
0&,& \;\;\;\;l<x.
\end{aligned}
\right.
\end{eqnarray}
The value of $g(x)$ can be easily seen from the contour in Fig.\ref{fig2}.

With these in hand, the integral turns out to be
\begin{eqnarray}
\int_{\mathcal{M}}\frac{e^{\frac{4\pi w}{\beta}}}{\left(e^{\frac{2\pi w}{\beta}}-e^{\frac{2\pi l}{\beta}}\right)^2\left(e^{\frac{2\pi w}{\beta}}-1\right)^2}=\frac{\beta  l \left(e^{\frac{2 \pi  l}{\beta }}+1\right)}{\left(e^{\frac{2 \pi  l}{\beta }}-1\right)^3},
\end{eqnarray}
and $\delta S(A)$ is simplified to
\begin{eqnarray}
\delta S(A)=-\frac{\mu\pi ^4 c^2 l   \coth \left(\frac{\pi  l}{\beta }\right)}{9 \beta ^3}.
\end{eqnarray}

\section{Correction of the R$\acute{\text{e}}$nyi entropy}
\subsection{Finite temperature}\label{renyit}
The leading order correction to $S_n(A)$ is given by (\ref{deltasn}). We already have $\left<T\overline{T}(w,\bar{w})\right>_{\mathcal{M}}$ and $\left<T\overline{T}(w,\bar{w})\right>_{\mathcal{M}^n}$. Previously we expand $\left<T\overline{T}(w,\bar{w})\right>_{\mathcal{M}^n}$ near $n=1$. Now we need its exact form, which is given by
\begin{eqnarray}
\left<T\overline{T}(w,\bar{w})\right>_{\mathcal{M}^n}=\left(\frac{c}{12}\right)^2\{z,w\}\{\bar{z},\bar{w}\},
\end{eqnarray}
where the Schwarzian derivatives $\{z,w\}$ and $\{\bar{z},\bar{w}\}$ are determined by the map
\begin{eqnarray}
w\to z=\left(\frac{e^{\frac{2\pi w}{\beta}}-1}{e^{\frac{2\pi w}{\beta}}-e^{\frac{2\pi l}{\beta}}}\right)^{\frac{1}{n}}.
\end{eqnarray}
Then $I\equiv\left<T\overline{T}\right>_{\mathcal{M}^n}-\left<T\overline{T}\right>_{\mathcal{M}}$ can be obtained. Using $w=x+i \tau$ and $\bar{w}=x-i \tau$, we arrive at
\begin{eqnarray}
\nonumber
\delta S_n(A)&=&\frac{-n\mu}{1-n}\int_{\mathcal{M}}I\\
&=&\frac{-n\mu}{1-n}\int_{-\infty}^{\infty} dx\int_0^{\beta} d\tau\;I(x,\tau).
\end{eqnarray}

We first do the $\tau$ integral, and the primitive function of $I(x,\tau)$ can be found. Let us call it $\mathcal{I}(x,\tau)$. As before there are two kinds of terms in $\mathcal{I}(x,\tau)$: the terms with $\log$ and the terms without $\log$. The analyses of them are similar with those in appendix \ref{integral}. After some carefull analyses we can express the integral as
\begin{eqnarray}\label{integralint}
\int_{-\infty}^{\infty} dx\int_0^{\beta} d\tau\;I(x,\tau)=\frac{\pi ^4 c^2}{36 \beta ^4}\left(\int_{-\infty}^0dx F(x)-\int_0^{\infty}dxF(x)+\int_{-\infty}^ldxG(x)-\int_l^{\infty}dx G(x)\right),
\end{eqnarray}
where
\begin{eqnarray}
F(x)&=& \frac{i \beta  \text{C}_1(x) \left(n^2-1\right) (-2\pi i)}{2 \pi  n^4 \left(e^{\frac{2 \pi  l}{\beta }}-1\right) \left(e^{\frac{4 \pi  x}{\beta }}-1\right)^3 \left(e^{\frac{2 \pi  l}{\beta }}-e^{\frac{4 \pi  x}{\beta }}\right)^3},\\
G(x)&=&\frac{-i \beta  \text{C}_2(x) \left(n^2-1\right) (-2\pi i)}{2 \pi  n^4 \left(e^{\frac{2 \pi  l}{\beta }}-1\right) \left(e^{\frac{2 \pi  l}{\beta }}-e^{\frac{4 \pi  x}{\beta }}\right)^3 \left(e^{\frac{4 \pi  l}{\beta }}-e^{\frac{4 \pi  x}{\beta }}\right)^3}.
\end{eqnarray}
Here $\text{C}_1, \text{C}_2$ are two functions of $x$, and their expressions are so long that we would not like to show them here. Notice that $F(x)$ has poles at $x=0,\;l/2$, and $G(x)$ has poles at $x=l/2,\;l$. With a cutoff $\epsilon$, the integral becomes
\begin{eqnarray}
\frac{\pi ^4 c^2}{36 \beta ^4}\left(\int_{-\infty}^{-\epsilon}dx F(x)-\int_{\epsilon}^{\frac{l}{2}-\epsilon}dxF(x)-\int_{\frac{l}{2}+\epsilon}^{\infty}dxF(x)+\int_{-\infty}^{\frac{l}{2}-\epsilon}dx G(x)+\int_{\frac{l}{2}+\epsilon}^{l-\epsilon}dxG(x)-\int_{l+\epsilon}^{\infty}dxG(x)\right).\;\;\;
\end{eqnarray}
The primitive functions of $F(x)$ and $G(x)$ can also be found, which are denoted by $\mathcal{F}(x),\;\mathcal{G}(x)$. Now the term in the bracket becomes
\begin{eqnarray}
\nonumber
&\;&\mathcal{F}(-\epsilon)+\mathcal{F}(\epsilon)+\mathcal{F}(\frac{l}{2}+\epsilon)-\mathcal{F}(\frac{l}{2}-\epsilon)-\mathcal{F}(\infty)-\mathcal{F}(-\infty)\\
&\;&+\mathcal{G}(l+\epsilon)+\mathcal{G}(l-\epsilon)+\mathcal{G}(\frac{l}{2}-\epsilon)-\mathcal{G}(\frac{l}{2}+\epsilon)-\mathcal{G}(\infty)-\mathcal{G}(-\infty).
\end{eqnarray}

We find that
\begin{eqnarray}
\mathcal{F}(\infty)+\mathcal{G}(\infty)&=&0,\\
\mathcal{F}(-\infty)+\mathcal{G}(-\infty)&=&0,\\
\mathcal{F}(\frac{l}{2}+\epsilon)-\mathcal{F}(\frac{l}{2}-\epsilon)&=&O(\epsilon),\\
\mathcal{G}(\frac{l}{2}-\epsilon)-\mathcal{G}(\frac{l}{2}+\epsilon)&=&O(\epsilon).
\end{eqnarray}
And
\begin{eqnarray}
\nonumber
\mathcal{F}(-\epsilon)+\mathcal{F}(\epsilon)+\mathcal{G}(l+\epsilon)+\mathcal{G}(l-\epsilon)&=&-\frac{2 \beta  l \left(n^2-1\right) \coth \left(\frac{\pi  l}{\beta }\right)}{n^2}+\frac{\beta ^4 \left(n^2-1\right)^2}{16 \pi ^3 n^4 \epsilon ^2}\\
\nonumber
&\;&-\frac{\beta ^2 \left(n^2-1\right)^2 \left(\cosh \left(\frac{2 \pi  l}{\beta }\right)-7\right) \text{csch}^2\left(\frac{\pi  l}{\beta }\right)}{24 \pi  n^4}\\
\nonumber
&\;&+\frac{\beta ^2 \left(n^2-1\right)^2 \coth ^2\left(\frac{\pi  l}{\beta }\right) \log \left(\frac{\beta  \sinh \left(\frac{\pi  l}{\beta }\right)}{2 (\pi  \epsilon )}\right)}{\pi  n^4}\\
&\;&+O(\epsilon^2).
\end{eqnarray}
Multiplying the prefactor back, we finally get
\begin{eqnarray}
\nonumber
\delta S_n(A)
&=&-\frac{\pi ^4 c^2 l \mu  (n+1) \coth \left(\frac{\pi  l}{\beta }\right)}{18 \beta ^3 n}+\frac{\pi  c^2 \mu  (n-1) (n+1)^2}{576 n^3 \epsilon ^2}\\
\nonumber
&\;&-\frac{\pi ^3 c^2 \mu  (n-1) (n+1)^2 \left(\cosh \left(\frac{2 \pi  l}{\beta }\right)-7\right) \text{csch}^2\left(\frac{\pi  l}{\beta }\right)}{864 \beta ^2 n^3}\\
\nonumber
&\;&+\frac{\pi ^3 c^2 \mu  (n-1) (n+1)^2 \coth ^2\left(\frac{\pi  l}{\beta }\right) \log \left(\frac{\beta  \sinh \left(\frac{\pi  l}{\beta }\right)}{2 \pi  \epsilon }\right)}{36 \beta ^2 n^3}\\
&\;&+O(\epsilon).
\end{eqnarray}

\subsection{Finite size}\label{renyil}
The discussion in this case is similar to the one in Appendix \ref{renyit}. Now we have
\begin{eqnarray}
\left<T\overline{T}(w,\bar{w})\right>_{\mathcal{M}^n}=\left(\frac{c}{12}\right)^2\{z,w\}\{\bar{z},\bar{w}\}
\end{eqnarray}
with the Schwarzian derivatives $\{z,w\}$ and $\{\bar{z},\bar{w}\}$ determined by the map
\begin{eqnarray}
w\to z=\left(\frac{\tan\left(\frac{\pi w}{L}\right)}{\tan\left(\frac{\pi w}{L}\right)-\tan \left(\frac{\pi  l}{L}\right)}\right)^{\frac{1}{n}}.
\end{eqnarray}
Defining $I\equiv\left<T\overline{T}\right>_{\mathcal{M}^n}-\left<T\overline{T}\right>_{\mathcal{M}}$ and using $w=x+i \tau,\;\bar{w}=x-i \tau$, we get
\begin{eqnarray}
\nonumber
\delta S_n(A)&=&\frac{-n\mu}{1-n}\int_{\mathcal{M}}I\\
&=&\frac{-n\mu}{1-n}\int_{-\infty}^{\infty} d\tau \int_0^{L} dx\;I(x,\tau).
\end{eqnarray}

Now we first do the $x$ integral, and the primitive function of $I(x,\tau)$ can be found which is denoted as $\mathcal{I}(x,\tau)$.  After some efforts we can express the integral as
\begin{eqnarray}\label{integralinl}
\int_{-\infty}^{\infty} d\tau\int_0^{L} dx\;I(x,\tau)=\frac{\pi ^4 c^2}{2304 L^4}\left(\int_{-\infty}^0d\tau F(\tau)-\int_0^{\infty}d\tau F(\tau)+\int_{-\infty}^0 d\tau G(\tau)-\int_0^{\infty}d\tau G(\tau)\right),\;
\end{eqnarray}
where
\begin{eqnarray}
F(\tau)&=& \frac{64 \text{D}_1(\tau) L \left(n^2-1\right)}{n^4 \left(-1+e^{\frac{2 i \pi  l}{L}}\right) \left(e^{\frac{4 \pi  \tau }{L}}-1\right)^3 \left(-e^{\frac{4 \pi  \tau }{L}}+e^{\frac{2 i \pi  l}{L}}\right)^3},\\
G(\tau)&=&\frac{64 \text{D}_2(\tau) L \left(n^2-1\right)}{n^4 \left(-1+e^{\frac{2 i \pi  l}{L}}\right) \left(e^{\frac{4 \pi  \tau }{L}}-1\right)^3 \left(-1+e^{\frac{4 \pi  \tau +2 i \pi  l}{L}}\right)^3}.
\end{eqnarray}
Here $\text{D}_1, \text{D}_2$ are two functions of $\tau$. We should notice that the form of the integral (\ref{integralinl}) is slightly different from the one of (\ref{integralint}), i.e. the $G$ integral is changed from $\int_{-\infty}^l-\int_l^{\infty}$ to $\int_{-\infty}^0-\int_0^{\infty}$. This difference is significant.

Now $F(\tau)$ only has a pole at $\tau=0$, so does $G(\tau)$. With a cutoff $\epsilon$,  the integral becomes
\begin{eqnarray}
\frac{\pi ^4 c^2}{2304 L^4}\left(\int_{-\infty}^{-\epsilon}d\tau F(\tau)-\int_{\epsilon}^{\infty}d\tau F(\tau)+\int_{-\infty}^{-\epsilon} d\tau G(\tau)-\int_{\epsilon}^{\infty}d\tau G(\tau)\right).\;\;\;
\end{eqnarray}
The primitive functions of $F(\tau)$ and $G(\tau)$ can be found out, which are denoted as $\mathcal{F}(\tau),\;\mathcal{G}(\tau)$. Now the term in the bracket becomes
\begin{eqnarray}
\nonumber
&\;&\mathcal{F}(\epsilon)+\mathcal{F}(-\epsilon)-\mathcal{F}(\infty)-\mathcal{F}(-\infty)+\mathcal{G}(\epsilon)+\mathcal{G}(-\epsilon)-\mathcal{G}(\infty)-\mathcal{G}(-\infty).
\end{eqnarray}

There are $\log$ terms in $\mathcal{F}(\tau)$ and $\mathcal{G}(\tau)$, so we should deal with them very carefully. After figuring out everything carefully, we finally get
\begin{eqnarray}
\nonumber
\delta S_n(A)&=&\frac{\pi  c^2 \mu  (n-1) (n+1)^2}{576 n^3 \epsilon ^2}\\
\nonumber
&\;&-\frac{\pi ^3 c^2 \mu  (n-1) (n+1)^2 \left(11 \cos \left(\frac{2 \pi  l}{L}\right)+19\right) \csc ^2\left(\frac{\pi  l}{L}\right)}{864 L^2 n^3}\\
\nonumber
&\;&+\frac{\pi ^3 c^2 \mu  (n-1) (n+1)^2 \cot ^2\left(\frac{\pi  l}{L}\right) \log \left(\frac{L \sin \left(\frac{\pi  l}{L}\right)}{2 \pi  \epsilon }\right)}{36 L^2 n^3}\\
&\;&+O(\epsilon^2).
\end{eqnarray}

\end{document}